\documentclass[twocolumn,prb,showpacs,preprintnumbers,amsmath,amssymb]{revtex4-1}

\usepackage{graphicx}
\usepackage{dcolumn}
\usepackage{bm}
\usepackage{tabularx}
\usepackage{multirow}
\usepackage{natbib}

\makeatletter
\AtBeginDocument{\@ifpackageloaded{natbib}{\ifNAT@numbers\if@filesw\immediate\write\@auxout{\string\global\string\NAT@numberstrue}\fi\fi}{}}
\makeatother
\begin{document}

\title {Topological surface states and Fermi arcs of the noncentrosymmetric Weyl semimetals TaAs, TaP, NbAs, and NbP}

\author{Yan Sun}
\affiliation{Max Planck Institute for Chemical Physics of Solids, 01187 Dresden, Germany}
\author{Shu-Chun Wu}
\affiliation{Max Planck Institute for Chemical Physics of Solids, 01187 Dresden, Germany}
\author{Binghai Yan}
\email{yan@cpfs.mpg.de}
\affiliation{Max Planck Institute for Chemical Physics of Solids, 01187 Dresden, Germany}
\affiliation{Max Planck Institute for the Physics of Complex Systems, 01187 Dresden,Germany}

\date{\today}

\begin{abstract}
Very recently the topological Weyl semimetal (WSM) state was predicted in the noncentrosymmetric compounds NbP, NbAs, TaP and TaAs 
and soon led to photoemission and transport experiments to verify the presumed topological properties such 
as Fermi arcs (unclosed Fermi surfaces) and the chiral anomaly.
In this work, we have performed fully \textit{ab initio} calculations of the surface band structures of these four WSM materials and
revealed the Fermi arcs with spin-momentum-locked spin texture.
On the (001) polar surface, the shape of the Fermi surface depends sensitively on the surface terminations (cations or anions), 
although they exhibit the same topology with arcs. The anion (P or As) terminated surfaces are found to fit 
recent photoemission measurements well. 
Such surface potential dependence indicates that the shape of the Fermi surface can be
sensitively manipulated by 
depositing guest species (such as K atoms), as we demonstrate.   
On the polar surface of a WSM without inversion symmetry, Rashba-type spin polarization naturally exists in the surface states and leads to
strong spin texture. By tracing the spin polarization of the Fermi surface, one can distinguish Fermi arcs from trivial Fermi circles. 
The four compounds NbP, NbAs, TaP, and TaAs present an increasing amplitude of spin-orbit coupling (SOC) in band structures.
By comparing their surface states, we reveal the evolution of topological Fermi arcs from the spin-degenerate Fermi circle 
to spin-split arcs when the SOC increases from zero to a finite value. 
Our work presents a comprehensive understanding of the topological surface states of WSMs, 
which will especially be helpful for future spin-revolved photoemission and transport experiments. 
\end{abstract}


\maketitle

\section{introduction}\label{intro}

The discovery of topological insulators (TIs) has refreshed our understanding of  
band theory and also promised new applications~\cite{qi2011RMP, Hasan:2010ku}. 
In a TI, the valence and conduction bands with opposite parity~\cite{Bernevig2006,fu2007} 
cross each other and open an energy gap at the band crossing point owing to spin-orbit coupling (SOC).  
As long as  time-reversal symmetry (TRS) is respected,  robust topological surface states appear 
inside the inverted bulk energy gap and usually exhibit a Dirac-cone-type energy dispersion with 
vortex-like spin texture [see Fig. 1(a)]. 
Interestingly, similar surface states were also predicted to exist on the surface of an exotic metal 
called Weyl semimetal (WSM)~\cite{Wan2011}.  
When the degeneracy is lifted by breaking the TRS or inversion symmetry, 
some band crossing points may remain gapless in a semimetal~\cite{Nielsen1981} and exhibit linear 
energy dispersions along all three-dimensional (3D) momenta ($k$) starting from this point, called the Weyl point, 
as an analog of 3D graphene. 
The Weyl point exhibits left- or right-hand chirality as a monopole of the Berry curvature (Chern flux) 
and always comes in pairs in the Brillouin zone (BZ). 
Between a pair of Weyl points with opposite chirality, robust edge states on the boundary are induced 
by the nonzero Chern number accumulated in the two-dimensional (2D) $k$ plane. 
However, no topological edge states appear in other regions because of the zero Chern number.
Therefore, unclosed Fermi lines formed by connecting these edge states,  called Fermi arcs, 
exist to connect the surface projection of two Weyl points with opposite parity [Fig. 1(b)]. 
The Fermi arc is apparently different from the Fermi surface of a TI or a trivial material, 
which is commonly a closed curve, and offers  strong evidence that a surface-sensitive 
technique such as angle-resolved photoemission spectroscopy (ARPES) can be used to identify the WSM. 
In addition, it should be noted that WSM is robust against any weak perturbation that 
preserves translational symmetry. For example, an exchange field or slight lattice 
distortion can only shift the positions of Weyl points. 
The Weyl points can only be annihilated in pairs of the opposite chirality.
WSMs were also predicted to exhibit exotic topological transport 
properties~\cite{Turner:2013tf,Hosur:2013eb,Vafek:2014hl,Parameswaran2014} (e.g., the chiral anomaly~\cite{Adler1969,Bell1969}) 
and offer promising applications for valleytronics and spintronics~\cite{Shekhar2015}.

Many WSM candidates have been predicted (e.g., the pyrochlore iridate Y$_{2}$Ir$_2$O$_7$~\cite{Wan2011} , 
HgCr$_2$Se$_4$~\cite{Xu:2011dy} , normal insulator (NI) and TI superlattices~\cite{Burkov:2011de}, 
solid solutions near the NI-TI phase transition point~\cite{murakami2007} with breaking TRS or inversion 
symmetry~\cite{Halasz2012} such as Hg$_{1-x-y}$Cd$_x$Mn$_y$Te~\cite{Bulmash2014}, LaBi$_{1-x}$Sb$_x$Te$_3$~\cite{Liu2014}, 
and TlBiS$_{1-x}$Se$_{2-x}$~\cite{Singh2012}, and Te and Se crystals under pressure~\cite{Hirayama2014}. 
However, none of the above candidates have been realized in experiment, except that possible transport 
evidence of negative magentoresistance was reported in Bi$_{0.97}$Sb$_{0.03}$~\cite{Kim2013}, 
ZrTe$_5$~\cite{Li2014ZrTe5}, and Na$_3$Bi~\cite{Xiong2015}.  Very recently, a new family of WSMs 
was predicted by band structure calculations~\cite{Weng:2014ue, Huang:2015uu} in the transition-metal monophosphides 
TaAs, TaP, NbAs, and NbP, in which the inversion symmetry is broken in their lattice while the TRS is still preserved. 
Topological Fermi arcs were soon observed in all four materials by ARPES~\cite{Lv2015TaAs, Xu2015TaAs,Yang:2015TaAs,Lv2015TaAsbulk, Liu2015NbPTaP,Xu2015NbAs,Xu:2015TaP}. 
Exciting transport phenomena were also reported, e.g., chiral magnetotransport in TaAs\cite{Huang2015anomaly,Zhang2015ABJ}, TaP~\cite{Shekhar:2015tp} and NbAs~\cite{Yang:2015vz},
extremely large magnetoresistance and high mobility in NbP~\cite{Shekhar:2015tp,Wang:2015wm} and NbAs~\cite{Ghimire2015NbAs}. 
This new family of WSMs presents complicated bulk and surface band structures. There are 12 pairs of Weyl points 
away from the high-symmetry lines in the BZ. Further, trivial surface states, topological surface states, and 
even bulk states overlap in energy and hybridize together on the surface~\cite{Lv2015TaAs,Xu2015TaAs, Xu2015NbAs,Liu2015NbPTaP}, 
which makes the identification of Fermi arcs difficult. Previous surface state calculations using the 
tight-binding method based on Wannier functions do not agree with recent ARPES Fermi surfaces, 
although they demonstrated the same topology~\cite{Weng:2014ue, Huang:2015uu}. Therefore, it is believed 
that fully \textit{\text{ab initio}} calculations, which take the local surface potential modification into account, are necessary to simulate the realistic surface states~\cite{Lv2015TaAs}. 

In this work, we have investigated the surface states of the WSM materials TaAs, TaP, NbAs, and NbP using \textit{ab initio} band structure calculations. We focus on the following questions: For a finite compound, what are the effects of surface atomic 
structures on the surface states and the topology of Fermi arcs? Which surface structure is the realistic 
one for an experiment? What is the spin texture of the Fermi arcs with respect to the existence of strong SOC and 
the Rashba effect? To compare all four compounds, what is the evolution of Fermi arcs given that different 
SOC strength in different materials may play a role? We organize this article in the following way: 
We first introduce our \textit{ab initio} method and the slab model with two types of possible atomic terminations, 
which is adopted to simulate the surface in Sec.~\ref{methods}. In Sec.~\ref{surfac}, we use TaP as an example to 
show the effect of different surface terminations and find that the cation-terminated surfaces agree with ARPES. 
In Sec.~\ref{Fermi-arcs}, we analyze the topology of Fermi surface and the spin textures to recognize the Fermi arcs. 
 In Sec.~\ref{Kdose}, we demonstrate that the shape of the Fermi surface (FS) can be sensitively tuned by surface 
modification such as  K deposition. Section \ref{conclusions} presents our conclusions. 

\begin{figure}[htbp]
\begin{center}
\includegraphics[width=0.45\textwidth]{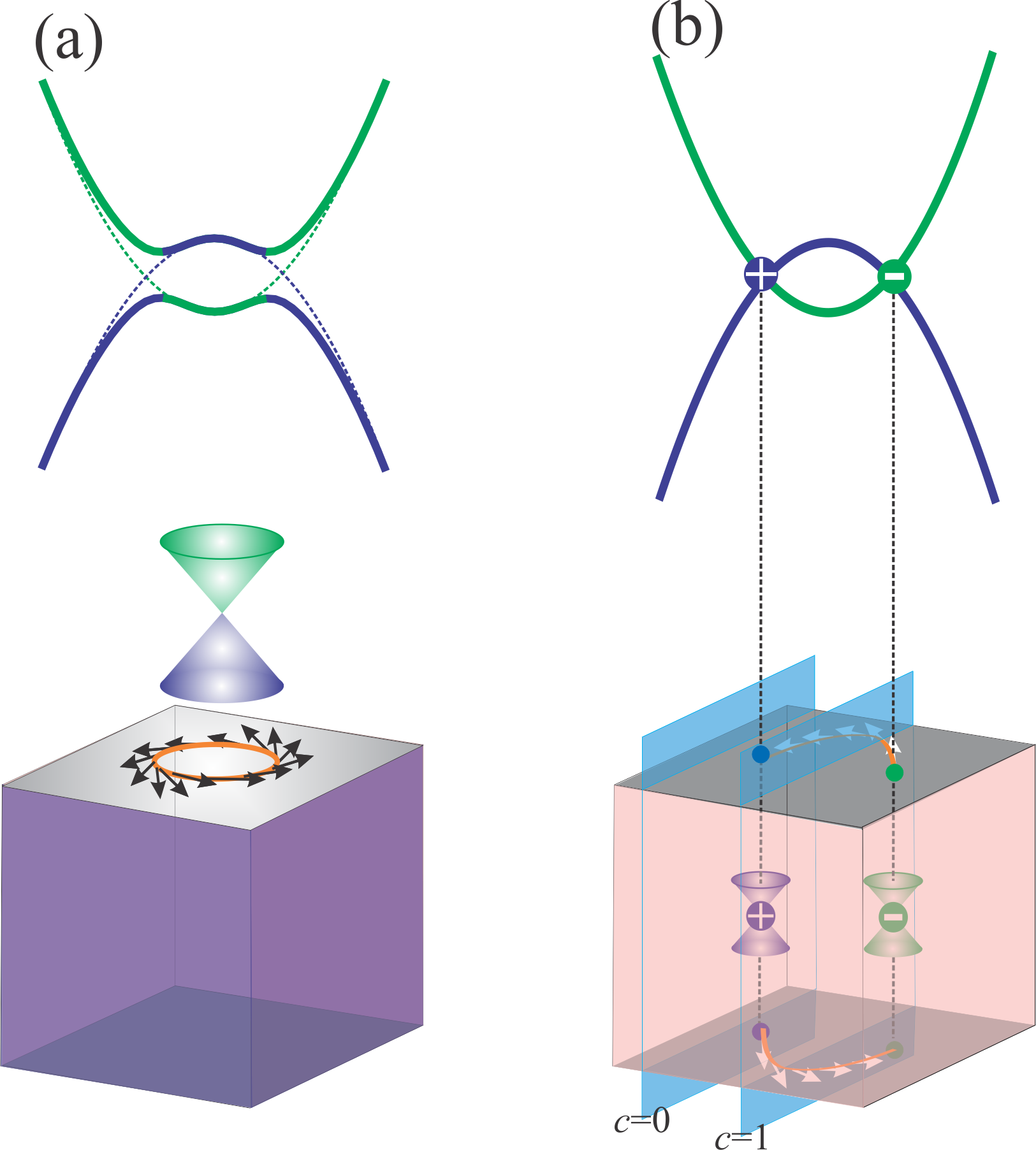}
\end{center}
\caption{
Schematics of the topological insulator and Weyl semimetal.
(a) A TI exhibits an energy gap with a band inversion. Topological surface states exhibits  Dirac-cone-type dispersion with spin texture.
(b) A WSM is gapless in the bulk and a pair of Weyl points (band crossing points) exists with opposite parity. 
Nonzero Chern number ($c$) only exists between Weyl points of the opposite chirality, which leads to
a spin-resolved surface Fermi arc.
}
\label{TI_WSM}
\end{figure}

\begin{figure}[htbp]
\begin{center}
\includegraphics[width=0.45\textwidth]{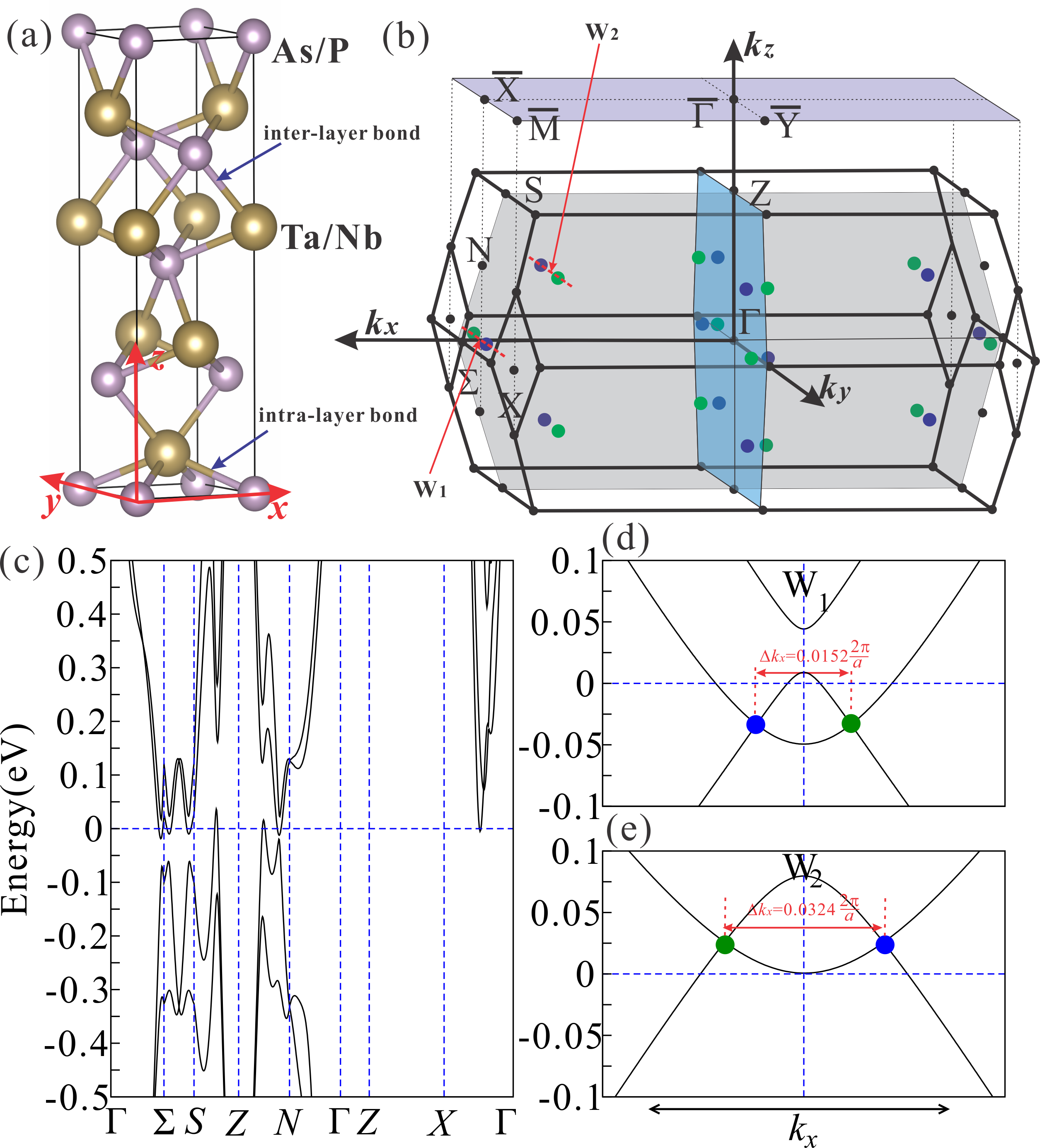}
\end{center}
\caption{
(a) Crystal structure of bulk TaAs, TaP NbAs, and NbP. The easy-cleaving planes are constructed by the inter-layer bonds.
(b) Bulk  BZ and projected surface BZ to the (001) plane. Twelve pairs of Weyl
points are denoted by green and blue dots to represent opposite chirality.
Weyl points in and out of the $k_z=0$ plane are denoted as W1
and W2, respectively. (c) Bulk band structures  of TaP along high-symmetry lines in
reciprocal space. (d) Local band structures of TaP along the connecting line between a pair of Weyl points. 
Corresponding paths are indicated in  (b) by red-dotted lines. The Fermi energy is shifted to zero.
}
\label{bulk_band}
\end{figure}

\begin{figure*}[htbp]
\begin{center}
\includegraphics[width=0.85\textwidth]{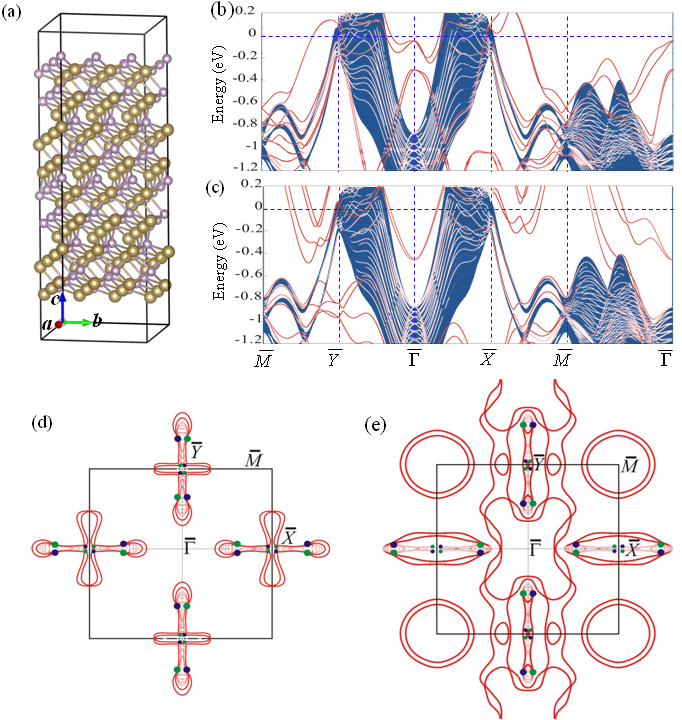}
\end{center}
\caption{
(a) Slab model along the $c$ direction for the surface state calculations.
Here we just show the thickness of 3 unit cells as an example.
In calculations, the slab with a thickness of  7 unit cells  was used.
(b) and (c) Surface band structures with P (b) and Ta (c) terminations along high-symmetry lines. Dark red colors represent the higher density projected
from the surface. Bulk projected states shown in blue  are also given as
the background. (d) and (e) Surface projected FSs with P (d) and Ta (e)
terminations. Weyl point projections in the 2D BZ are marked as blue and
green dots.
}
\label{slab band}
\end{figure*}

\section{methods}\label{methods}

Density-functional theory (DFT) calculations are performed with the projected augmented wave (PAW) potential  implemented in the Vienna
$Ab~initio$ Simulation Package (\textsc{vasp})~\cite{kresse1996, kresse1996-ms}.
The exchange-correlation energy is considered within the generalized gradient approximation (GGA)~\cite{perdew1996}, and the energy cutoff was set to
300 eV for a plane wave basis.

The four compounds share the same face-centered tetragonal lattice (space group $I4_1 md$, No. 109) without inversion symmetry, 
as presented in Fig. 2(a).
Since the (001) surface is the cleaved surface in ARPES measurements~\cite{Lv2015TaAs, Xu2015TaAs,Yang:2015TaAs, Liu2015NbPTaP,Xu2015NbAs,Xu:2015TaP}, we choose this surface in our study here.
The (001) surface is a polar surface terminated with either cation (Ta or Nb) or anion (As or P) atoms, where dangling bonds appear.
Although the compound is three dimensional, there are still some easy-cleaving planes to form a surface, as indicated in Fig. 3(a). 
One only needs to break two Ta-P bonds in one unit cell in this easy plane, whereas one has to break four bonds otherwise. 
This simple observation is further confirmed with our total energy calculations, in which the formation energy of the 
easy plane is 2.5 eV per unit cell lower than that of the other plane.
Therefore we construct a slab model by cutting the bulk at the easy-cleaving planes. 
As illustrated in Fig.~\ref{slab band}, the top and bottom surfaces are terminated with P and Ta atoms, respectively.
We find that the dipole effect resulting from asymmetric surface terminations has a negligible effect on the band structure, 
because the slab itself is a semimetal that easily screens the weak dipole field. 
The slab is 7 unit cells thick and is periodic in the $xy$ plane but separated by a 10 \AA\ vacuum space 
in the $z$ direction. 
Our test with thicker slabs (e.g., 9 and 11 unit cells thick) gives the same surface band structure and Fermi surfaces as the 7-unit-cell one. 
A dense k-point grid of $400\times400$ is adopted to calculate the 2D Fermi surfaces. 
To distinguish surface states easily from bulk states, 
we add the weighted  wave function contributions from the outermost unit cell (eight atomic layers) in the band structure and Fermi surface graphics.

\begin{figure*}[htbp]
\begin{center}
\includegraphics[width=0.85\textwidth]{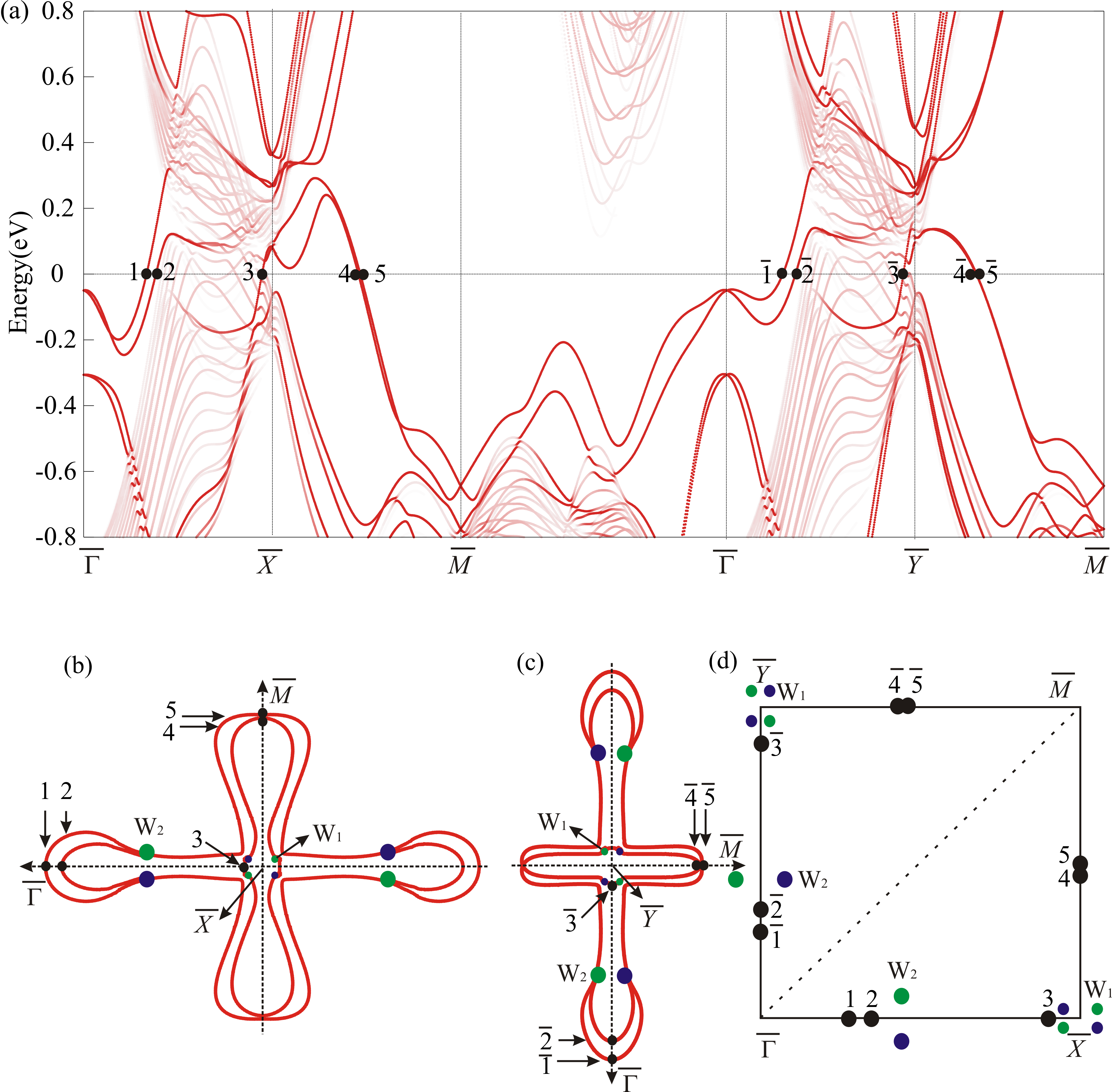}
\end{center}
\caption{
(a) Surface projected band structure with P terminations along the high-symmetry lines of 
$\overline{\Gamma}$-${\overline X}$-${\overline M}$-$\overline{\Gamma}$-${\overline Y}$-${\overline M}$.
(b) and (c) Surface projected  FSs around (a) ${\overline X}$ and (b) ${\overline Y}$ points.
(d) Schematics showing the Fermi crossing along the
loops of $\overline{\Gamma}$-${\overline X}$-${\overline M}$-$\overline{\Gamma}$
and $\overline{\Gamma}$-${\overline Y}$-${\overline M}$-$\overline{\Gamma}$.
Black dots are the crossing points, while blue and green dots represent Weyl points
with opposite chiralities.
}
\label{slab fs}
\end{figure*}

\section{Results}\label{results}

\subsection{Bulk band structure}\label{bulk}

Taking TaP as an example, we see that, because of the lack of inversion symmetry, the double spin degeneracy splits the band structure,
as shown in Fig. 2(c). Consistent with recent calculations ~\cite{Weng:2014ue, Huang:2015uu}, 
there are a total of 12 pairs of Weyl points in the whole BZ, as presented in Fig. 2(b):
4 in the $k_{z}=0$ plane and the other 8 pairs in a plane close to $k_{z}=\pm\pi/c$.
Any pair of Weyl points with opposite chiralities are connected by the mirror symmetry with respect
to the mirror planes of $k_x=0$ or $k_y=0$. For convenience, we denote the Weyl points in the $k_z=0$ 
plane as Weyl point 1 (W1), and the others are denoted as Weyl point 2 (W2). 
From the local energy dispersions around Weyl points one can find that the splits of the two types of 
Weyl points in $k$ space are very different. As shown in Figs. 2(d) and 2(e), the distance between  nearby W2 points in $k$ space is almost twice that between nearby W1 points. 
Moreover, the two types of Weyl points do not lie at the same energy level, with W2 being  $\sim$60 meV 
above W1. The Fermi level is located between the two types of Weal
points,  $\sim$50 meV above W$_1$ and $\sim$10 meV below W2. Therefore, W2 points lie much  closer
to the Fermi energy (the charge neutral point) than W1 for TaP and also for other three compounds.
This indicates that the W2 points may be more significant than W1 to affect the experiment observation of the chiral anomaly effect in magneto-transport, which agrees with recent transport experiments~\cite{Huang2015anomaly,Zhang2015ABJ,Wang:2015wm,Yang:2015vz, Shekhar:2015tp}.

\subsection{Surface band structure}\label{surfac}

As shown in Fig. 3(a), after cutting the bulk at the easy-cleaving bond, 
the symmetry of the slab is reduced from $C_{4v}$ to $C_{2v}$. As presented in Figs. 3(b) and 3(c), 
with $C_{4v}$ symmetry, the bulk bands give the same energy dispersions along 
$\overline {M}$-$\overline {Y}$-$\overline{\Gamma}$ and
$\overline {M}$-$\overline {X}$-$\overline{\Gamma}$.
However, the surface bands along the two directions are different owing to the 
reduced $C_{2v}$ symmetry. And for the surface energy dispersion, the projected 2D 
FSs around $\overline {X}$ and $\overline {Y}$ points are also different for both P and 
Ta terminations, as shown in Figs. 3(d) and 3(e).

Comparing surface band structures in Figs. 3(b) and 3(c), we can see that the surface 
bands behave very differently for the two types of terminations. 
Between $\overline {M}$-$\overline {Y}$, there are four electron-type surface 
bands crossing the Fermi level for the Ta-terminated surface, whereas only two 
holelike bands cut the Fermi level for P termination. Moreover, compared to the
strong splitting of the four electron surface bands in the Ta-terminated surface, 
the two holelike surface bands are nearly spin degenerated for the P-terminated state.
Such differences can be also seen in the  $\overline {M}$-$\overline {X}$ direction.
Another obvious difference in the surface band structure appears in the    
$\overline {M}$-$\overline{\Gamma}$ direction. In this direction, the surface bands around the Fermi
energy are all below the Fermi level, exhibiting as valences band for the P-terminated surface.
In the Ta termination, the surface bands extend across a larger energy window around the Fermi energy,  cutting the 
Fermi level four times.
For both terminations, most of the surface bands are located in the bulk band gaps.
However, there are still several surface bands submerged into bulk states and 
strong bulk-surface hybridizations exists, especially in the zone around the $\overline {X}$
and $\overline {Y}$ points. Because of the strong hybridization, some surface bands exhibit 
as broken pieces, being  no longer continuous curves, such as the energy dispersion along 
$\overline {Y}$-$\overline{\Gamma}$-$\overline {X}$ in the P termination.

\begin{figure*}[htbp]
\begin{center}
\includegraphics[width=0.85\textwidth]{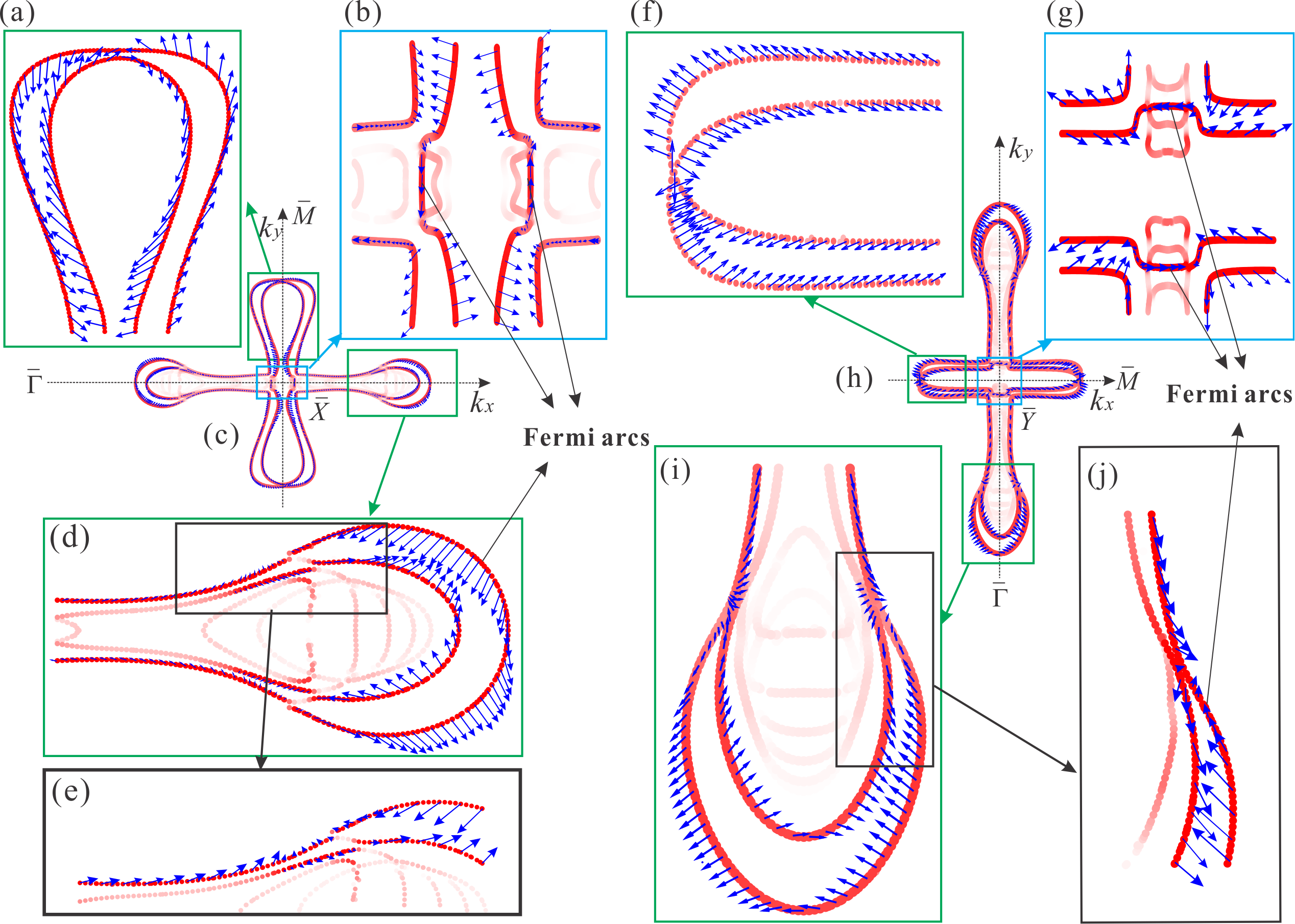}
\end{center}
\caption{
Spin textures of the surface FS. Dark red colors represent the
surface FS and blue arrows are spin orientations.
}
\label{spin_textrue}
\end{figure*}

Surface projected FSs with P and Ta terminations at the Fermi
level are shown in Figs. 3(d) and 3(e). Consistent with corresponding
surface band structures, the two terminations give very different FSs, and 
the FSs in the Ta-terminated surface are much more complicated than those of the P termination.
For the Ta-terminated side, the surface projected FS spreads all over the 2D BZ, whereas the
P-terminated FS is mainly  located around the $\overline {X}$ and $\overline {Y}$
points. The P-terminated surface gives spoon-shaped FSs around the projected
W2 along both $\overline{\Gamma}$-$\overline {X}$ and $\overline{\Gamma}$-$\overline {Y}$
directions, whereas there are too many FSs around the projected W2 points for the 
Ta-terminated surface. Comparing the shape and topology of the FSs projected from the two
different terminations, we can see that the cation(e.g., P)-terminated surface fits
the ARPES measured results much better~\cite{Lv2015TaAs, Xu2015TaAs,Yang:2015TaAs, Liu2015NbPTaP,Xu2015NbAs}.
Therefore, our detailed analysis will be focused on the P-terminated surface in the following. 
In addition, we note that a recent measurement on TaP possibly present surface states that can be attributed to the Ta-terminated surface~\cite{Xu:2015TaP}.

\subsection{Fermi arcs}\label{Fermi-arcs}

The complexity of the surface state originates from fundamental properties
of surface and bulk electronic structures around the Fermi level. 
At the Fermi energy, normal bulk electron and hole pockets coexist with Weyl points, as shown in Figs. 2(c)--2(e).
This is further confirmed by our recent Fermi surface measurement by quantum oscillations~\cite{Shekhar:2015tp}.
From the bulk band structures in Figs. 2(c)--2(e) we have seen that the Weyl points
are not exactly located at the Fermi level; the electron and hole pockets crossing the Fermi level are very
close to the Weyl points in $k$ space, especially for W1. Therefore the hybridization between
bulk and surface states cannot be avoided. Another reason for the complexity
of the surface state originates from the dangling bonds. 
As we mentioned above, dangling bonds appear on both Ta and P terminations. 
They will present surface states with strong Rashba splitting, and one needs to distinguish
them from topologically nontrivial surface states. Further, the anion (Ta) or cation (P) terminated surfaces exhibit strong
surface band bending in the outermost surface atomic layers, which will shift the surface state in energy. 
Moreover, Fermi arcs from W1 and W2 may also hybridize together.

Because of these reasons, we need to make a detailed analysis of the surface
Fermi arcs step by step in the following. First, we confirm the most
fundamental characteristic of the WSM surface state, the existence of Fermi arcs.
A simple and effective way to identify the nontrivial surface state in WSMs
is to count the number of FS crossings through a generic closed loop
in the 2D BZ~\cite{Weng:2014ue}. Here we choose the loop of
$\overline{\Gamma}$-$\overline {X}$-$\overline {M}$-$\overline{\Gamma}$ and
$\overline{\Gamma}$-$\overline {Y}$-$\overline {M}$-$\overline{\Gamma}$ to count
the number of crossings.
Although  FSs from dangling bonds with strong Rashba splitting exist on the surface,
they are closed Fermi circles, and they always give even numbers of Fermi crossings for a closed loop.
In contrast, Fermi surfaces related to Weyl points are nonclosed arcs ~\cite{Wan2011}
and will contribute odd numbers of FS crossings.
On the (001) surface two W2 points with the same chirality are projected to
the same point in the (001) 2D BZ; hence both chosen loops enclose three Weyl points,
one W1 and two doubly degenerate W2, where W1 and W2 have opposite chirality. 
Therefore the total crossing number is expected to be odd.

\begin{figure*}[htbp]
\begin{center}
\includegraphics[width=0.85\textwidth]{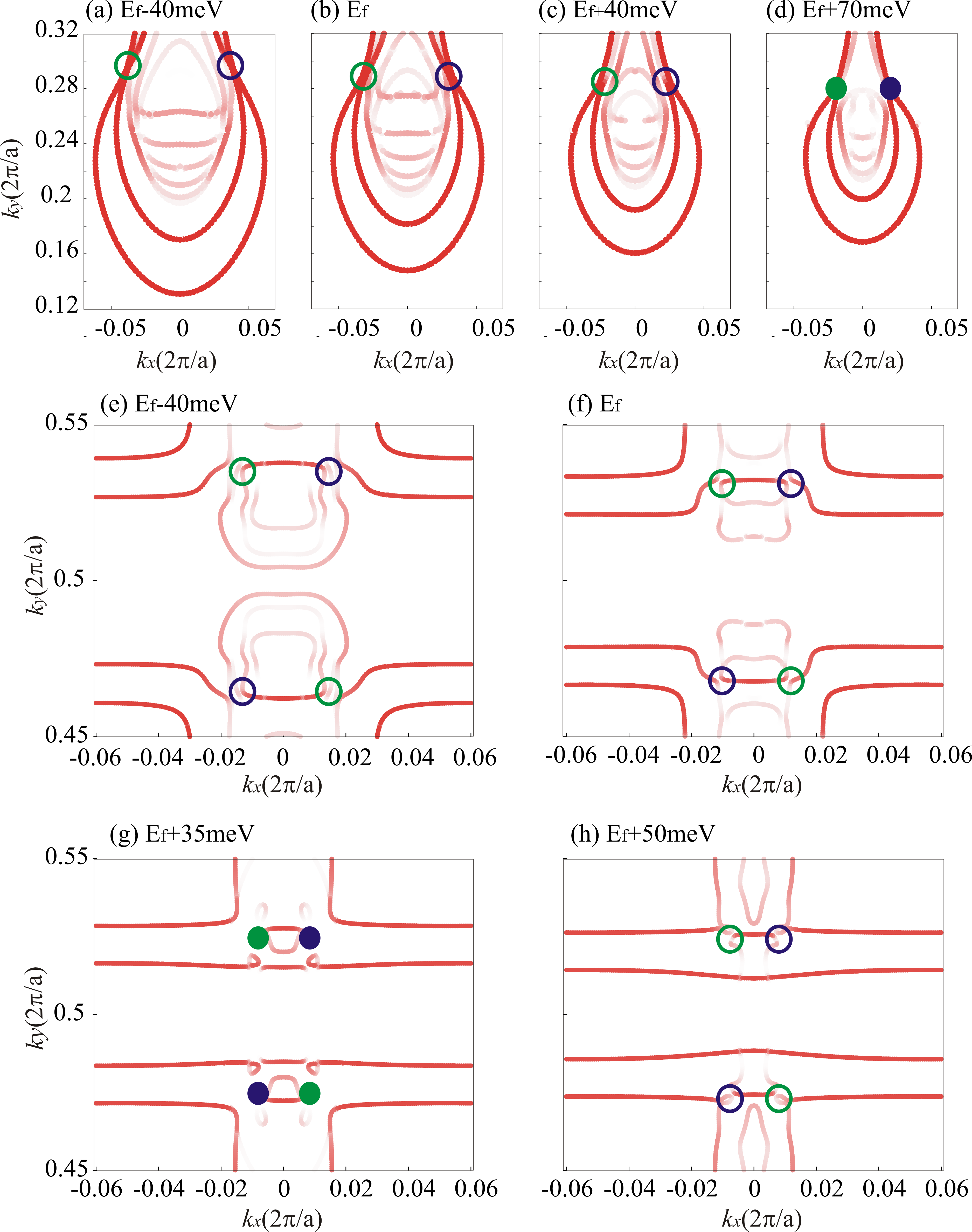}
\end{center}
\caption{
Evolution of the local Fermi surfaces around (a)--(d) W2 and (e)--(h) W1 in
the P-terminated surface with varying  Fermi energy. Green and blue circles
are the terminations of surface Fermi arcs. Green and blue dots are the exact
2D projections from bulk Weyl points.
}
\label{fs evolution with e }
\end{figure*}

The surface projected band structure with P termination along the high-symmetry lines of
$\overline{\Gamma}$-$\overline {X}$-$\overline {M}$-$\overline{\Gamma}$-$\overline {Y}$-$\overline {M}$
is given in Fig. 4(a). Since there are no  surface bands cutting the Fermi energy between
$\overline{\Gamma}$-$\overline {M}$, we just need to consider the crossing point along
$\overline{\Gamma}$-$\overline {X}$-$\overline {M}$ and
$\overline{\Gamma}$-$\overline {Y}$-$\overline {M}$. From Fig. 4(a) we can see
that the crossing number along each closed loop is five, an odd number as we expected.
Consistent results were also seen for the surface projected FSs. As shown in
Figs. 4(b) and 4(d), the closed loop
$\overline{\Gamma}$-$\overline {X}$-$\overline {M}$-$\overline{\Gamma}$
encloses two doubly degenerate positive W2 and one negative W1 with opposite
chiralities.
Along $\overline{\Gamma}$-$\overline {X}$, the loop crosses the FS three times
(twice around W2 and once around W1), and the other two crossings appear between
$\overline {X}$-$\overline {M}$ at two nearly degenerate  FSs. So the total number is  
also five, as obtained from the surface band structure. A similar result can be also observed 
from the FS around the $\overline {Y}$ point in Figs. 4(c) and 4(d). Therefore, the nontrivial surface
state of WSMs is verified by the odd number of Fermi crossings from  both band structures and FSs.

\begin{figure*}[htbp]
\begin{center}
\includegraphics[width=0.85\textwidth]{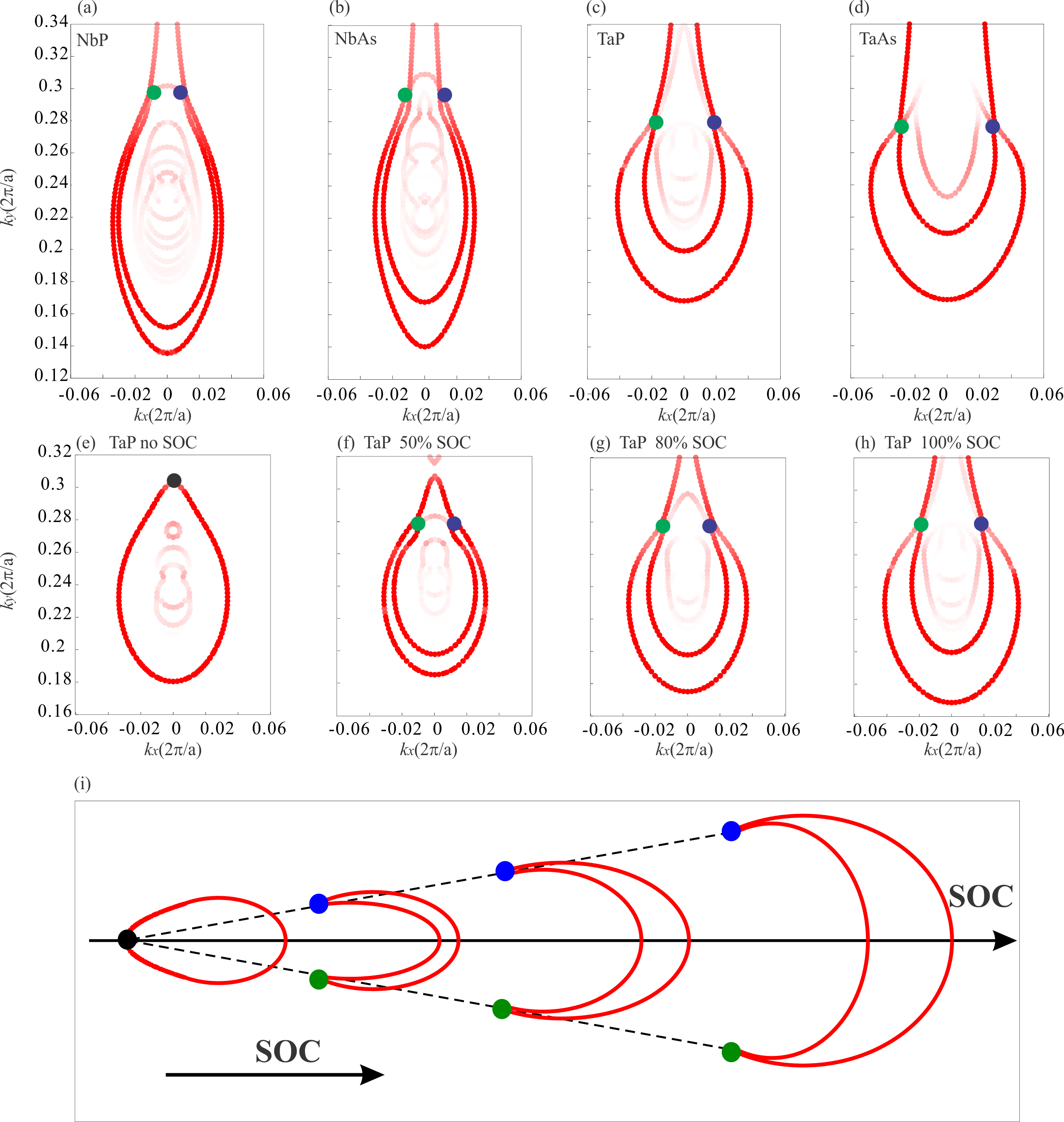}
\end{center}
\caption{
(a)--(d) Comparison of the FSs around W2 for the four compounds 
NbP, NbAs, TaP, and TaAs.
(e)--(h) The split increases with increasing  SOC strength.
(i) Schematic diagram of Fermi arcs with increasing  SOC strength.
}
\label{fs_vary_soc*}
\end{figure*}

Although we can confirm the nontrivial surface state from the odd number  of crossings
between the closed 2D loop and the FS, the details of the Fermi arcs remains unclear. From the overview in Figs. 4(b) and 4(c) we can see that the surface projected
FSs around both $\overline {X}$ and $\overline {Y}$ seem to consist of two 
closed Fermi circles and two nonclosed Fermi arcs. As shown
in Fig. 4(b), along $\overline{\Gamma}$-${\overline X}$ around
W2, there are two pieces on the left side of W2, whereas only one is present on the right side. Therefore, one FS must
be the Fermi arc terminated at one pair of positive and negative W2 points.
A similar Fermi arc also exists around the $\overline {Y}$ point [as presented in
Fig. 4(c)], which is below the W2 in the  
$\overline{\Gamma}$-$\overline {Y}$ direction.

However, we still do not know which pieces of the FSs are the real Fermi arcs.
Moreover, because of the strong hybridization between surface and bulk states, 
as noted in the surface band structure shown in Fig. 4(a), it is not 
easy to distinguish surface and bulk bands. To solve these problems,
we analyzed the spin textures of the surface projected FSs.
The spin textures for the projected 2D FS around $\overline {X}$ and $\overline {Y}$ 
points are shown on the left and right sides in Fig. 4(a), respectively. 
From the overview of the spin textures, one can find that the spin
moments are almost zero for bulk projected FSs. Therefore, we can easily
distinguish the surface FSs from the bulk state by the magnitude of the spin moment.

In addition, the direction of spin moments can help us to identify the Fermi
arcs from trivial FSs. In the $\overline{\Gamma}$-$\overline {X}$ direction,
two FSs on the left side of W2 have opposite spin textures, the inner FS
being right-handed while the outer one being left handed, as illustrated in the enlarged view of the
local FSs in Fig. 5(d). Comparing the left and right sides of W2, we can see
that the inner FS on the right side has a continuous spin texture with the
left FS, as presented in the further enlarged view of the local FSs in Fig. 5(e). Therefore,
the outer FS is a clear Fermi arc connecting to a  W2 pair with opposite chiralities.

Compared to the Fermi arcs terminated at W2, the  W1-related 
Fermi arcs are relatively shorter, as we have seen in the bulk energy 
dispersion [see Figs. 2(d) and 2(e)].
In the $\overline {X}$-$\overline {M}$ direction, the loop crosses the FS twice
around the middle of $\overline {X}$-$\overline {M}$. These two pieces of 
FSs have opposite spin textures around the crossing point, as shown in Fig. 5(a). 
However, as the FSs approach the $\overline {X}$ point, the spin directions become discontinuous. As presented in the enlarged view of the local FSs and spin textures 
in Fig. 5(b), the spin directions along the inner FS are not smoothly 
connected. For example, both the lower and 
upper pieces of the left inner FS   have $+k_{y}$ components, whereas the $s_y$ component for the 
middle piece is along $-k_{y}$, and the spin amplitudes are reduced to zero 
around the transition point. Therefore, the seemingly continuous FSs belong to 
different wave functions. Since the Weyl point W1 is located just  near 
the spin transition point, the middle piece of the FSs may be understood as the Fermi 
arcs connecting one pair of negative and positive W1.

The FSs around the $\overline {Y}$ point show similar features. The outer FS below 
W2 is the Fermi arc terminated at one pair of W2. A spin direction
transition also exists for the $s_x$ component around $\overline {Y}$ points
along the $k_x$ direction, implying  the existence of Fermi arcs connecting to W1. 
Therefore, the Fermi arcs are identified by the combination of surface projected 
FS and spin textures.

We must stress that the above understanding of Fermi arcs is only for the purpose of 
simple interpretation of complicated FSs. Given the complicated hybridization between trivial FSs and arcs, 
the most valid way to determine the topology is still counting the FS crossings as discussed above.

Since each Fermi arc  ends at one pair of positive and negative Weyl points on
the projected (001) 2D BZ, after the identification of Fermi arcs, we can confirm 
the location of projected Weyl points from the surface FS. However, we found that 
the positions of the two ends of Fermi arcs in the 2D BZ are not exactly equal to the bulk Weyl point projection. 
Taking W2 as the example, from the bulk band structures we have found that  W2 in the 3D BZ is
$(\pm0.016\frac{2\pi}{a},\pm0.273\frac{2\pi}{a},\pm0.294\frac{2\pi}{c'})$,
where $c'=c/2$, $a=3.318 \;\rm\AA,$ and $c=11.363\;\rm \AA$ ~\cite{Willerstrom1983}, but the coordinates of the two ends of the surface Fermi arcs are
$(\pm0.026\frac{2\pi}{a},\pm0.297\frac{2\pi}{a})$, which is not equal to the 2D projection from the bulk W2.

The nonequal coordinates between the bulk projected Weyl point and the Fermi arc
terminations is a direct consequence of the slab model employed. Since the anion P$^{3-}$ 
and cation Ta$^{3+}$ have opposite charges, the (001) slab model constructs 
pole surfaces, and the P termination has a slightly higher electric potential. Consequently,  bands
near the surface with P and Ta terminations will be shifted upward and downward, respectively.
Therefore, we expect to find  accurate Weyl points projections at an energy slightly
above the Fermi energy for the P-terminated surface. We note that the effect of the surface
potential  modifying the energy of the Fermi arcs was also proposed recently~\cite{Li:2015wp}.
Figures 6(a)--6(d) give the evolution of the 2D FS around W2 as a result of the
upshifting of energy. The split between the two ends of the Fermi arcs decreases as
the energy shifts upwardly. At an energy of 70 meV above the Fermi energy we
get the coordinates of the Fermi arc terminations  closest to the bulk W2 projection, 
as presented in Fig. 6(d). Though the coordinates of the Fermi arc terminations
are not exactly equal to the bulk Weyl point projection, the existence of
Fermi arcs is robust as the shifting of energy occurs across a wide energy window of $E_f \pm 100$ meV. 
Applying the same method to the FSs around $\overline {Y}$, 
as shown in Figs. 6(e)--6(h), we found that, when energy shifts upward to
35 meV above the Fermi level, the W1 coordinates derived from the termination of
Fermi arcs in the 2D BZ is  closest to the bulk projection. So we found both W1 and W2 
projections in the P-terminated surface by shifting the Fermi level, 
which should be taken into account to analyze the ARPES surface states.

\subsection{Comparison of NbP, NbAs, TaP, and TaAs}\label{compare}

Using  the same method, we also confirmed Fermi arcs in the other three Weyl
semimetals (NbP, NbAs, and  TaAs).
Because of their same crystal symmetry and bulk electronic band orders, the four compounds
give similar Fermi arcs in the P- and As-terminated (001) surface. Though the FSs in all
four compounds give the same topological feature of Fermi arcs, there are still many differences.
Compared to W1, the $k$-space splitting is much larger and the bulk hybridization
is relatively weaker around W2; therefore, it is better to take the Fermi arcs terminated at one
pair of W2 as an example to show the similarities and differences among the four WSMs.

In Figs. 7(a)--7(d), we present the 2D FSs around W2 for the four WSMs with
P and As terminations. All  four compounds give spoon-shaped FSs in the
$\overline{\Gamma}$-$\overline {Y}$ direction. As was the case for TaP, for NbP, NbAs, and TaAs, there are two pieces of FSs
below the pair of W2 in the direction of $\overline{\Gamma}$-$\overline {Y}$,
and only one piece remains above  W2.
Therefore, one of the two pieces of FSs must be a Fermi arc. Through the trace of spin textures
as we performed in TaP, we found that the inner FSs below W2 have the same spin orientation
as the FSs above W2, as shown in Figs. 7(a), 7(b), and 7(d). Hence the outer pieces of FSs below W2
are apparently the nonclosed Fermi arcs, which is just same as the  TaP case.
The split strength between positive and negative W2 
increases  in the order  of NbP, NbAs, TaP, and TaAs. Meanwhile,
one can see that the narrow, long spoon-shaped Fermi arcs smoothly change to wide and short.
Since the bands around the Fermi energy are mainly Ta and Nb \textit{d} orbitals dominated by small
As and P \textit{p}-orbital hybridizations~\cite{Weng:2014ue}, and the strength of SOC effect is proportional
to the weight of elements, the crucial physical mechanism for the evolution of Fermi arcs in NbP, NbAs, TaP, and TaAs
is the increased SOC amplitude.

\begin{figure*}[htbp]
\begin{center}
\includegraphics[width=0.85\textwidth]{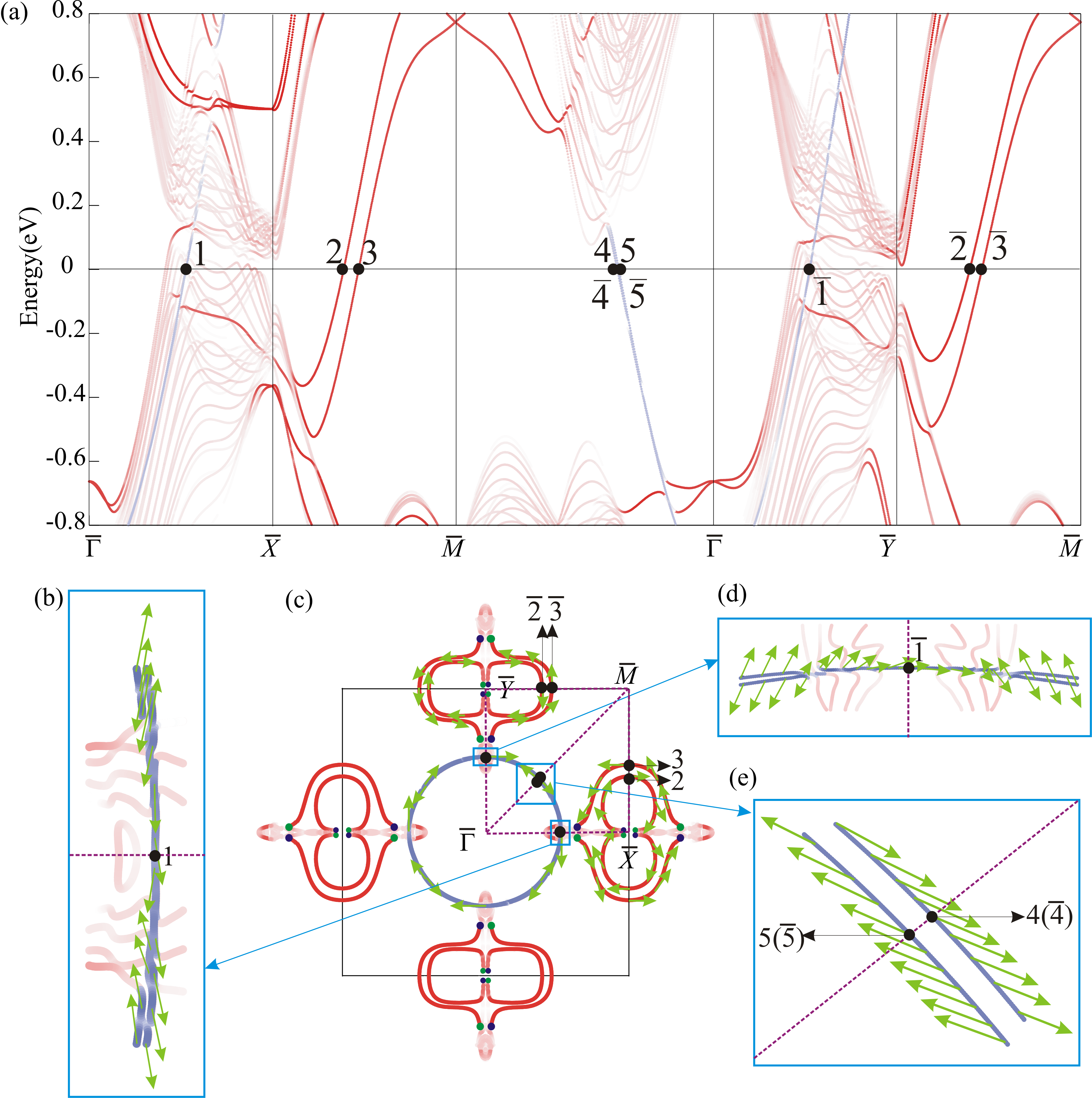}
\end{center}
\caption{
Surface energy dispersion (a) and FSs (b)--(e) of TaP with one atomic layer of K deposited on the P-terminated 
surface. Red curves represent the projected bands and FSs from the top  unit cell (eight atomic layers) of TaP.
Bands and FSs in  blue are projected from the deposited single K atomic layer. Green arrows are the spin orientations.
Black dots are the crossing points, while blue and green dots represent Weyl points
with opposite chiralities. Fermi arcs are denoted as 1, 2, and 3 and $\overline{1}$, $\overline{2}$, and $\overline{3}$.  
}
\label{K-dosing}
\end{figure*}

To validate this scenario, we analyzed the projected FSs of TaP with
increasing SOC strength from zero to a finite value, while keeping all other parameters fixed.
As shown in Fig. 7(e), without the inclusion of the SOC effect, the surface FS is
just a trivial spin-degenerate Fermi circle. As long as the SOC effect is taken
into consideration, the single Fermi circle splits into two pieces. Different
from Rashba splitting, there is only one closed Fermi circle
in the lifted FS,  the other one being a nonclosed Fermi arc terminated at one
pair of negative and positive W2. This process implies the existence of band
inversion even without the SOC effect, and SOC just gives rise to the Weyl points,
which is consistent with previous bulk electronic structure analysis~\cite{Weng:2014ue}.
As presented in Figs. 7(f)--7(h), increasing the SOC strength also expands the split between negative and positive W2, 
as the evolutions in the WSM compounds by the ordering of  NbP, NbAs, TaP, and TaAs.
Therefore, the splitting strength between one pair of Weyl points in WSMs with respect to the mirror plane
are mainly determined by their SOC strength. As a general feature in this family of WSMs, we give a
schematic diagram for the evolution of the Fermi arcs with increasing SOC in Fig. 7(i). Without the inclusion of the SOC effect, the
FS exhibits as a closed Fermi circle, symmetric with respect to the  $k_x$ (or $k_y$) axis. As long as
the SOC effect is taken into consideration, the spin-degenerate point on the $k_x$ (or $k_y$) axis is lifted into one pair of
Weyl points with respect to the $k_x$ (or $k_y$) axis. Meanwhile, the Fermi circle is transformed into a
nonclosed Fermi arc terminated at two Weyl points. With increasing SOC strength, the splitting between positive and
negative Weyl points  also increases.

\subsection{Surface modification} \label{Kdose}

Like the surface and edge states in topological insulators and
quantum Hall insulators, the nontrivial surface state in Weyl
semimetals is also protected by the topology of bulk band order. Therefore,
the existence of surface topological Fermi arcs in Weyl semimetals should be robust against 
weak surface perturbations. Nonetheless, the detailed shapes of the Fermi surface can surely be
modified by the surface perturbations, similar to the case of  TI surfaces~\cite{Shu-Chun:2014,Binghai-review:2012}
To check the robustness of the Fermi arc and detect the manipulation of
surface Fermi arcs, we calculated the FSs with potassium adsorption on the P-terminated surface.

We deposited one K layer on top of the P-terminated surface of TaP.
After the K-dosing, the surface band structure is dramatically changed,
and the most important difference is the disappearance of the surface bands
along $\overline{\Gamma}$-$\overline {X}$ and $\overline{\Gamma}$-$\overline {Y}$, as shown
in Figs. 4(a) and 8(a). Besides the change of bands originated from the TaP surface,
a pair of new bands from the K-$s$ orbital appears around the Fermi level.
As illustrated in Fig. 8(c), the K-$s$ orbital follows
a nearly regular circle around the  $\overline{\Gamma}$ point in the projected 2D FS.
Though the shape of the FSs change dramatically compared to the pristine  FSs,
the topological feature does not change.

Compared to the pristine surface, the Fermi arcs in the K-dosed surface are relatively simple,
as can be easily identified even without the help of spin texture.
As shown in Fig. 8(c), there are
two types of Fermi arcs in both $k_x$ and $k_y$ directions, one terminated at two
W1 points and  the other terminated at two W2 points with opposite chirality.
In the pristine surface, one pair of positive and negative Weyl points connected by a Fermi
arc has mirror symmetry with respect to $k_x=0$ (or $k_y=0$), whereas, in the K-dosed surface,
the mirror plane for each Fermi arc is transformed to $k_y=\pm\pi/a$ (or $k_x=\pm\pi/a$).
Because of this transformation, the Fermi arcs become much longer and easier to detect.
More important is that the Fermi arcs connecting to W1 become more
apparent owing to the weak bulk-surface hybridization in the  $\overline {X}$-$\overline{M}$
and  $\overline {Y}$-$\overline {M}$  directions, as shown by the band structure and FS 
in Figs. 8(a) and 8(c).

Although the Fermi arcs have been identified, from self-consistent point of view,
it is also necessary to further confirm the nontrivial surface state by the  crossing number between FSs and closed loops.
Choosing the same loop as in the pristine surface, we found that the crossing number between the
loop of $\overline{\Gamma}$-$\overline {X}$-$\overline {M}$-$\overline{\Gamma}$
($\overline{\Gamma}$-$\overline {X}$-$\overline {M}$-$\overline{\Gamma}$ ) is still odd (5 times),
which implies a nontrivial Fermi arc surface state.
Because of the strong surface-bulk hybridization, as shown in Figs. 8(b) and 8(d),
only one piece of the FS is preserved for the K $s$-orbital-constructed FSs near the line of
$\overline{\Gamma}$-$\overline {X}$ and $\overline{\Gamma}$-$\overline {X}$, which leads
to one crossing between the loop and the FS in each of these two directions. In both the $\overline {X}$-$\overline {M}$
and $\overline {Y}$-$\overline {M}$ directions, there are two Fermi arcs cutting the loop.
Between $\overline {M}$-$\overline{\Gamma}$, only the K $s$-orbital-constructed FSs cut the loop.
From the enlarged view of the local FSs in Fig. 8(e) we can see that the FSs in this direction consist of two bands with spin up and spin down. Though the SOC parameter for the $s$ orbital is zero,
the coupling of the K $s$ orbital with the Ta $d$ and P $p$ orbitals opens a small gap for the K $s$ bands.
Hence, the FSs cut the loop twice between $\overline {M}$-$\overline{\Gamma}$. Therefore, each
of the chosen closed loops crosses the FS five times, an odd number as expected.
In ARPES experiments, K-dosing on the surface can be easily done. We predict that the above FS modification
might be detectable. However, the K-related Fermi circles could be absent in an FS measurement if the K layer is not well ordered.

\section{Conclusions}\label{conclusions}

In conclusion, by using first-principles calculations we have systematically studied the surface Fermi arcs
in the family of noncentrosymmetric Weyl semimetals comprising NbP, NbAs, TaP, and TaAs.
In two types of terminations along the (001) direction, the P- and As-terminated surfaces fit the ARPES measurements better.
From the odd number of Fermi crossings in a generic loop we can confirm the existence of Fermi arcs on the surface FSs.
The continuous spin textures of the FSs can help us to distinguish the real Fermi arcs from trivial surface bands.
The evolution of the Fermi arcs in  NbP, NbAs, TaP, and TaAs is mainly due to the
increased SOC strength. Through K-dosing we found that the shape of the Fermi arcs can be manipulated by varying the
surface conditions, although the Fermi-arc-characterized surface state is robust to surface perturbations.
These results will be helpful for a clear understanding of the surface Fermi arcs in the TaP type of Weyl semimetals.

\begin{acknowledgments}
We are grateful for Z.-K. Liu, L.-X. Yang, X. Dai and Y.-L. Chen for helpful discussions. This work  was financially supported  by the  Deutsche Forschungsgemeinschaft DFG  (Project No. EB 518/1-1 of DFG-SPP 1666 "Topological Insulators", and SFB 1143) 
and by the ERC (Advanced Grant No. 291472 "Idea Heusler").  
\end{acknowledgments}

\bibliographystyle{ieeetr}

\bibliography{references}

\end{document}